%% file: lpep++.tex
\documentclass[conference]{IEEEtran}

\usepackage[T1]{fontenc}
\usepackage[utf8]{inputenc}

\usepackage{csquotes}
\usepackage[english]{babel}

\usepackage{bm}
\usepackage{amsmath}
\usepackage{amssymb}
\usepackage{amsfonts}
\usepackage{algorithmic}

\usepackage{balance}
\usepackage{textcomp}
\usepackage{microtype}

\usepackage[style=ieee, backend=bibtex]{biblatex}
\usepackage[acronym,hyperfirst=false,nohypertypes={acronym}]{glossaries}
\usepackage{cleveref}

\usepackage{xcolor}
\usepackage{graphicx}
\usepackage{mathtools}
\usepackage{booktabs}

\graphicspath{{./figures}}

\renewcommand\gcd{\mathrm{gcd}}
\newcommand{\mathvf}[1]{\ensuremath{\bm{#1}}} %

\newcommand{\tildet}{\tilde{t}}
\newcommand{\tildeP}{\tilde{P}}

\def\BibTeX{{\rm B\kern-.05em{\sc i\kern-.025em b}\kern-.08em
    T\kern-.1667em\lower.7ex\hbox{E}\kern-.125emX}}

\loadglsentries[main]{glossary.inc.tex}

\begin{document}

\title{Robust and Deterministic Scheduling\\
  of Power Grid Actors}

\author{
  \IEEEauthorblockN{Emilie Frost, Eric MSP Veith and Lars Fischer}
  \IEEEauthorblockA{OFFIS e.V.\\
    Oldenburg, Germany \\
    Email: \texttt{firstname.lastname@offis.de}}
}

\maketitle

\begin{abstract}
  Modern power grids need to cope with increasingly decentralized, volatile
  energy sources as well as new business models such as virtual power plants
  constituted from battery swarms. This warrants both, day-ahead planning of
  larger schedules for power plants, as well as short-term contracting to
  counter forecast deviations or to accommodate dynamics of the intra-day
  markets. In addition, the geographic distribution of renewable energy
  sources forces scheduling algorithms with a hugely different communication
  link qualities. In this paper, we present an extension to the \gls{LPEP}, dubbed
  \emph{\gls{LPEP}++}. It draws on the strength of the \gls{LPEP} to find the
  optimal solution of the combinatorial power demand-supply problem with
  string guarantees in acceptable time and extends it with facilities for
  long-term planning, parallel negotiations and reduces its memory footprint.
  We furthermore show its robustness towards volatile communication link
  quality.

\end{abstract}


\glsresetall

\section{Introduction}

Globally, there has been a huge shift set in motion to provide a decarbonized
energy supply. These goals have been widely propagated through many agreements
in different countries; the European Union's goal to be climate-neutral by
2050 might serve as an example to this claim~\cite{EuropeanCommission2018}.
The reduction of CO\textsubscript{2} emissions is, in many countries, achieved
by increasing the share of volatile, i.e., weather-dependant renewable energy
sources such as wind and \gls{PV}.

The shift from a centrally managed, hierarchically-structured power grid
towards with power plants feeding into the transmission grid towards
decentralized power generation using renewables was one of the reasons that
has given rise to the concept of the smart grid, as it requires the
introduction of extensive \gls{ICT} infrastructure to coordinate demand and
supply. Arguably, this decentralized power generation also calls for a
decentralized, \emph{divide-et-impera}-style
approach~\cite{veith2017universal}. This means that a \gls{MAS} is used to
manage power demand and supply---or other aspects of power grid operations,
such as the provision of reactive power---, integrating a huge share of
renewable-energy-based generators.

This new design of the power grid needs \gls{ICT} connectivity to deliver on
its promise of an efficient power supply. However, wind parks and \gls{PV}
plants are usually erected at locations that are sensible from the perspective
of harvesting the wind's or sun's power, but not with regards to an seamless
inclusion into the grid's \gls{ICT} network. In addition, newer prosumer and
market concepts, such as \glspl{VPP} and battery swarm storage, are
increasingly constituted from
\gls{MAS}~\cite{troschel2009towards,niesse2012market,wang2011intelligent,ebell2018coordinated,dai2019optimal}.

Many of these approaches abstract from the communication medium, assuming
losless link quality with negligible delay. The research project \emph{LarGo!}
considers as one of the first research projects the complex interaction
between the power grid and its \gls{ICT} infrastructure. Mainly focused on
software rollouts, a resilient power grid scheduling is one of the research
questions of the project~\cite{kintzler2018large}.

In the scope of the research question, this paper describes an extension to
the \gls{LPEP}. We show the resiliency of the \gls{LPEP} with regards to an
impaired communication medium, i.e., specifically to (intermittently) high
delays. The procotol's and the underlying demand-supply
solver's ability to draft a new schedule for power provisioning or consumption
in non-optimal \gls{ICT} situations is shown. We also present an extension to
the protocol, named \emph{\gls{LPEP}++} that allows a more efficient
convergence towards complete schedules.

The remainder of this paper is structured as follows: We will introduce
related work in \Cref{sec:related-work}. As an extension, we provide a
description of the current state of the \gls{LPEP} in
\Cref{sec:lpep-fundamentals}. Based on this, \Cref{sec:protocol-extension}
details the modifications and forms the main contribution of this paper. In
\Cref{sec:testing}, we provide experimental results to substantiate our claims
towards \gls{LPEP} improvement with results obtained from simulation. We
conclude in \Cref{sec:conclusion}, where we also provide an outlook for future
work.

\section{Related Work}
\label{sec:related-work}

One of the ancestral behavioral protocols for \glspl{MAS} is the \emph{Contract
Net Protcol} by \textcite{smith1980contract}. Here, agents announce tasks using
broadcast messages for other agents to bid on. The announcement also contains
the ranking process, i.e., bids delivered by other agents are ranked according
to metrics such as estimated time to task completion. The announcer, or task
manager, then awards the task to a specific node, informing all other nodes in
the process. The awarded node can then additionally choose to break the task
up into smaller subtasks and sub-contract them through a similar procedure.

The general broadcast-bidding-awarding structure of behavior laid down in the
\emph{Contract Net Procol} has influenced many (negotiation) protocols for
distributed computation. In many cases, additional ideas are brought in to add
efficiency, to speed up the negotiation, or to reduce the amount of messages
or data being sent. The
\gls{LPEP}~\cite{veith2013lightweight,veith2017universal} specifies initial
messages (requests for or offers of power) as broadcasts, but models the
overlay networks the agents use on the power grid in which the agents'
physical entities represent, imposing rules on message routing that limit
message propagation, introducing the concept of dynamic neighborhoods where
supply and demand have as little physical line meter between them as possible,
reducing the line loss. Responses are routed directly through a dynamic
routing table on each node that is being built during the request stage.

Additionally, \textcite{Shen1995} worked towards eschewing the initial
broadcast stage. They employ multicasting---i.e., the network protocol
concept~\cite{rfc1112}---for the task announcement messages, creating interest
groups to which agents can subscribe. \textcite{Wanyama2007} reduce the
number of negotiation rounds until consensus is reached, limiting the scope of
agent coalitions to a \emph{group-choice problem} and basing their negotiation
approach on game theory, replacing explicit knowledge through message
exchanges by implicit knowledge coming from a game-theoretic model of the
negotiation process. \textcite{Garcia2017} have reduced the number of messages
per negotiation, assuming a control theory problem behind the agents'
communication and implementing an asynchronous, event-based protocol based on
a discretized model that is decoupled from the state of the agent's neighbors.

The aforementioned publication by \textcite{Olfati-saber2007} also emphasizes
the effectiveness of neighborhood concepts, based on \emph{small-world
networks} by \textcite{watts1998collective}---being one of the hallmark works
on overlay topologies for distributed computing---, and referring to the
weightings introduced by \textcite{xiao2004fast}. The two works heavily
influenced the later, much-celebrated \emph{small-world model} for \glspl{MAS}
by \textcite{olfati2005ultrafast}. The \gls{COHDA} protocol by
\textcite{hinrichs2013cohda}---the key competitor to \gls{LPEP}++---builds on the small-world model;
\textcite{niesse2017local} also note that fast convergence or the quantitative
guarantee of convergence do not necessarily mean that the optimal solution to
a problem is found, but that the \gls{ICT} overlay network topology influences
the search for a solution with certain \gls{MAS} protocols.

In the context of a \gls{CPS}, fully decentralized \gls{MAS} approaches to a
problem can be viewed with suspicion. After all, there is no way to control or
``look into'' the process as it happens. The statement of the convergence
problem by \textcite{hanachi2004protocol} mentioned above is approached by the
authors through a \emph{protocol moderator}, i.e., an explicit middleman.
Similarly, for \gls{COHDA}, \textcite{niesse2016controlled} propose an
observer-controller architecture for the in its core completely
decentralized protocol. The questions these approaches rise is whether how
certain behavior can be formulated as being expected, rather than just
exhibited. It is expressed in the move from specifications to \emph{contracts}
in component design.

The \gls{LPEP} features a different approach towards contracting. It leverages
the power of \glspl{TVL} to model demand and supply. Thus, it is guaranteed to
arrive at the optimal solution for a given set of input
data~\parencite{veith2017agent}. Its current shortcoming is its granularity: The
\gls{LPEP} starts negotiation for every timestep anew. This allows for a
maximum of flexibility, but has high costs in terms of efficiency, especially
if a complete schedule needs to be reconsidered due to intraday flexibilities.

\section{LPEP Fundamentals}
\label{sec:lpep-fundamentals}

\subsection{Communication Protocol}

The \gls{LPEP} as it exists now~\cite{veith2017universal} has, at its core,
the so-called \emph{Four-Way Handshake}. Agents initiate a negotiation with
either a \emph{Demand Notication} or \emph{Offer Notification}, depending on
whether they request additional power or offer it. To understand both
cases, one most take the concept of the \emph{power equilibrium} into account.
This constitutes the state in which power demand and supply match. Whenever an
agent detects a deviation from the state of equilibrium, e.g., based on a
node-local forecast, it initiates the negotiation. When the disequilibrium is
caused by a surplus of power, a \emph{Offer Notification} starts the
negotiation; consequently, the \emph{Demand Notification} expresses the
agent's request for power from other nodes.

These initial messages are then routed through an overlay network modelled
according to the underlying power grid infrastructure. The rules governing
the message exchange let message propagation boundaries form, depending on the
contribution of neighboring node (called ``match-or-forward'' rule). This
favors power equilibria. During this stage of the negotiation that essentially
employs broadcasting, the ad-hoc routing table for the directly-routed answers
is built. The routing metric is the impedance of the local power line, i.e.,
the goal is to reduce line loss.

When other agents answer, they send the matching \emph{Offer Notification} to
an initial \emph{Demand Notification}, and vice versa. The difference is the
presence of an explicit message ID in the answer, which is the ID of the
initial request. When the initiating agent receives answers, it can start its
internal solver to match its request with the replies it received; this is
discussed in \Cref{ssec:lpep-fundamentals/solver}. When the solver finds a
solution, all agents whose offer are taken receive an \emph{Acceptance
Notification} and, in turn, reply with an \emph{Acceptance Acknowledgement
Notification}. This fourth step allows agents to withdraw replies, e.g., when
forecasts deviate.

\begin{figure}
  \includegraphics[width=\linewidth]{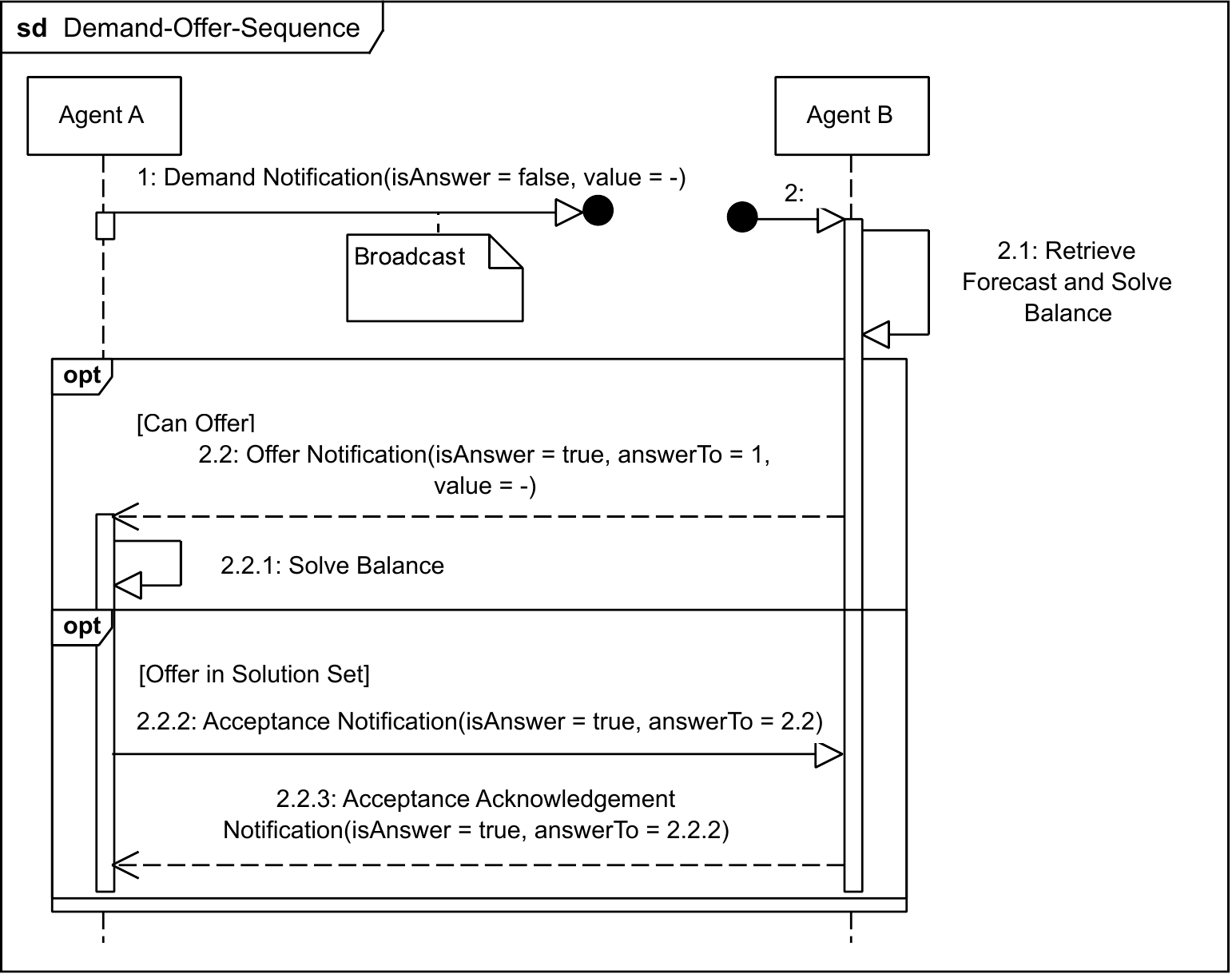}
  \caption{The Four-Way Handshake}
  \label{fig:way}
\end{figure}

The Four-Way Handshake is depicted in \Cref{fig:way}. For a
more detailed, extensive discussion, please refer to
\cite{veith2017universal}.

\subsection{Solver}
\label{ssec:lpep-fundamentals/solver}

Any demand or offer is mathematically described as a mapping \(\tilde{t}
\mapsto P\), where \(\tilde{t} = [t_1; t_2]\) is a time interval for which
\(P\) is valid. The vector of all absolute power values is \(\mathbf{P} =
(|P_0|, |P_1|, |P_2|, \ldots)\); the vector of all time interval lengths is
\(\mathbf{\tilde{t}} = (t_{0,2} - t_{0,1}, t_{1,2} - t_{1,1}, t_{2,2} -
t_{2,1}, \ldots)\).

To convert this multi-valued optimization problem into a Boolean-valued one,
we use the greatest common divisor to formulate each demand and offer in terms
of atoms sized

\begin{equation} \label{gcd}
  \Delta P = \gcd(\mathvf{P}), \; \Delta t = \gcd(\mathvf{\tilde{t}})~,
\end{equation}

\noindent each representing one part of it:

\begin{equation}\label{eq:o-simple}
  x_{i, \tildet, \tildeP} = \begin{cases}
    1 & \parbox[t]{0.65\linewidth}{ if the agent \(i\) influences the power
      grid in the time subinterval \(\tilde{t}\) with power from the power
      subinterval \(\tilde{P}\),}\\
    0 & \text{otherwise.}
  \end{cases}
\end{equation}

Using these atoms, the \emph{requirements function} for each demand or offer
expresses complete, potentially partial acceptance, or decline:

\begin{equation}
  \mathrm{r}_i(\mathvf{x}_{i,\tildet,\tildeP}) = \begin{cases}
    1 & \parbox[t]{0.6\linewidth}{if \(\mathvf{x}_{i,\tildet,\tildeP}\)
      denotes a valid interval for accepting the requirement
      from agent \(i\),}\\
    0 & \text{otherwise.}
  \end{cases}
\end{equation}

The simplest case that is always present is ``accept fully, or do not accept
at all'':

\begin{equation}
  \mathrm{r}_i(\mathvf{x}_{i,\tildet,\tildeP}) =
    \bigwedge_i \mathvf{x}_{i,\tildet,\tildeP} \vee
    \bigwedge_i \mathvf{\bar{x}}_{i,\tildet,\tildeP}
\end{equation}

Next, symmetric functions for each time subinterval are used to model all
possible arrangements of the atoms:

\begin{equation}
  \mathrm{S}^{\frac{P_0}{\Delta P}}_{k}(\mathvf{x}_{i,\tildet,\tildeP}) =
    \begin{cases}
      1 & \parbox[t]{0.5\linewidth}{if \(n\) variables in
        \(\mathvf{x}_{i,\tildet,\tildeP}\) are 1,}\\
      0 & otherwise,
    \end{cases}
\end{equation}

\noindent with \(k = 1, 2, \ldots, \frac{|\tilde{t}|}{\Delta t}\).

In order to arrive at an solution, the agent must determine the exact cover:

\begin{equation}
  \mathrm{C}(\mathvf{x}_{i,\tildet,\tildeP}) = \bigwedge_k
    \mathrm{S}^{\frac{P_0}{\Delta P}}_k(\mathvf{x}_{i,\tildet = k,\tildeP})
    \bigwedge_i \mathrm{r}_i(\mathvf{x}_{i,\tildet,\tildeP})~.
\end{equation}

A representation using \glspl{TVL}, implemented using the XBOOLE system,
enables an efficient calculation of this
cover~\cite{steinbach1992xboole,veith2017universal}.

\section{Protocol Extension}
\label{sec:protocol-extension}
In the originial \gls{LPEP}, with every new input, the agent with the disequilibrium tries to solve it. In the \gls{LPEP}++, if not the complete disequilibrium could be solved, the agent after a certain time determines the power which can be afforded. This might be relevant
for use cases like trading on energy markets, where agents negotiate about trades which
cannot be fulfilled anymore. Therefore, as much power as possible needs to be afforded
in order to keep overstretching of existing commitments as low as possible.
Since the agent with the disequilibrium does not have knowledge of all agents of the network, it may not wait until it received every offer or demand. A timeout was implemented for the conversion. The inquiring agent then starts a timer when sending out the initial \emph{Offer} or\emph{Request Notification}. The timer being expired, the agent determines the power which could be afforded based on the available. After solving the disequilibrium as best as possible, it sends out \emph{Acceptance Notifications} and the agents follow the \emph{Four-Way Handshake} of the \gls{LPEP}. 
The number of seconds the timer expires depends on the size of the network. It needs to be high enough so that the \emph{Offer} or \emph{Demand Notifications} of as many agents as possible have already arrived in order to not exclude any power. In a network of six agents, the timer was set to 0.02 seconds.

Another possible extension is considering a total replanning of schedules with the \gls{LPEP}.
If an imbalance occurs in more than one interval of the schedule, it is required to
replan every relevant interval of the schedule. Therefore, the agent determining the
disequilibrium sends out an \emph{Offer} or \emph{Demand Notification} with the missing power and time
mapping for the whole timeframe which is concerned.

Agents receiving the call for supply may answer with a time-power-mapping in any kind of time interval length. If supplying power is for example only possible in parts of the
requested time, the agent only replies with power for this time and sends its reply as time-power-mapping for the possible time.

\begin{figure} 
  \includegraphics[width=\linewidth]{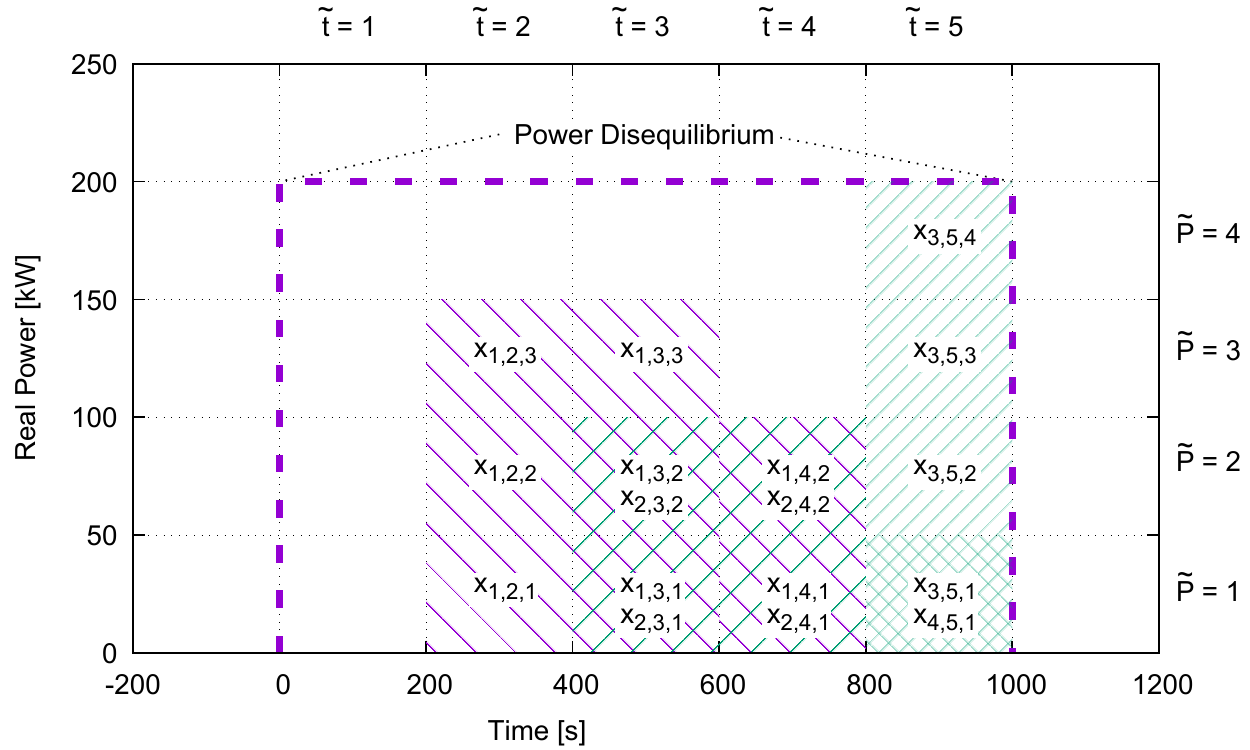}
  \caption{Boolean demand and supply}
  \label{fig:boolean}
\end{figure}

The agent with the disequilibrium receiving the answers, divides them by building atoms for each time interval given by the other agents.
Since the solver describes demands or offers mathematically as a mapping from time to power \(\tilde{t}
\mapsto P\), where \(\tilde{t} = [t_1; t_2]\) is a time interval for which power value
\(P\) is valid, multiple time intervals may be considered.
\Cref{fig:boolean} shows an example of the power balance state after the discretization. In the example, the acceptance function is shown from \cref{req}.

\begin{equation}
{X}_{1} = x_{1,2,1} \land x_{1,2,2} \land x_{1,2,3}
\end{equation}

\begin{equation}
{X}_{2} = x_{1,3,1} \land x_{1,3,2} \land x_{1,3,3}
\end{equation}

\begin{equation}
{X}_{3} = x_{1,4,1} \land x_{1,4,2} \land x_{1,4,3}
\end{equation}

\begin{align}
\label{req}
\mathrm{r}_1(\mathvf{x}_{1,\tildet,\tildeP}) =
\mathvf\bar{{X}}_{1} \land \mathvf\bar{{X}}_{2} \land \mathvf\bar{{X}}_{3}
\\ \lor {X}_{1} \land \mathvf\bar{{X}}_{2} \land \mathvf\bar{{X}}_{3}
\\ \lor {X}_{1} \land {X}_{2} \land \mathvf\bar{{X}}_{3}
\\ \lor {X}_{1} \land {X}_{2} \land {X}_{3}
\end{align}

The \gls{LPEP}++ then does not intitiate a new negotiation or each timestep, but it allows to consider several timesteps in one negotiation and so consider entire schedules in one negotiation. Thus, the \gls{LPEP}++ keeps the maximum of the flexibility and furthermore improves effiency.

\section{Solver Modification}
\label{sec:solver-modification}
Due to heterogenous power offers, sizing each atom of the power of the agents by using the greatest common divisor leads to many different variables. As an alternative \cite{veith2017universal} states using an interval partioning algorithm to lower the number of variables.
Therefore, the solvers way to divide the \emph{Offer} or \emph{Demand Notification} and the disequilibrium needs to be modificated. It determines interval margins of the
power values and uses these interval margins to convert the problem into the boolean domain. Each entry in the TVL for each offer or demand is an atom which states whether the agent influences the power grid for the given power subinterval. The dividing of the notifications into subintervals here is depending on the offers and demands of all other agents and the disequilibrium.
Considering power intervals \(\mathbf{P} = (|P_0|, |P_1|, |P_2|, \ldots)\) formed by all power values available for the solver, atoms are represented as:
\begin{equation}\label{eq:interval}
  x_{i, \tildet, \tildeP} = \begin{cases}
    1 & \parbox[t]{0.65\linewidth}{ if the power value of agent \(i\) is within
    current power subinterval \(\tilde{P}\) and influences power grid at 
    time subinterval  \(\mathvf{x}_{i,\tildet,\tildeP}\),}\\
    0 & \text{otherwise.}
  \end{cases}
\end{equation}
Taking the power values \(\mathbf{P} = (5, 51, 150)\) as an example, thus according to~\cref{gcd} 

\begin{equation} \label{gcd-ex}
  \Delta (5, 51, 150) = \gcd(5, 51, 150)
\end{equation}

results in atoms sized

\begin{equation} \label{gcd-ex-2}
  \gcd(5, 51, 150) = 1
\end{equation}

To determine the number of atoms \(\mathbf{N}\)\textsubscript{gcd} for the sizing of the atoms with the gcd, divide the maximal power value by the size of atoms.

\begin{equation} \label{gcd-ex-3}
  N_{gcd} = \frac{P_{max}}{\Delta P}
\end{equation}

\begin{equation} \label{gcd-ex-4}
  N_{gcd} = 150
\end{equation}
Thus, with sizing the atoms according to the gcd, the solver considers 150 atoms.
Taking into account sizing the atoms with the interval partitioning algorithm, the number of atoms \(\mathbf{N}\)\textsubscript{intervals} equals the number of intervals. According to power values \(\mathbf{P} = (5, 51, 150)\), three intervals are existing \(P_{intervals} = [0; 5], [6;51], [52;150]\).

\begin{equation} \label{gcd-ex-5}
  N_{intervals} = 3
\end{equation}

The example shows the reduction of the number of variables by using the interval paritioning algorithm. 

\section{Simulation-based Testing}
\label{sec:testing}
\subsection{\gls{LPEP}++ compared to COHDA}
The \gls{LPEP} guarantees the optimal solution to be found ~\parencite{veith2017agent}. \gls{COHDA}, on the other hand, does not give this guarantee. In order to contrast further behaviours of the two systems, a setting was implemented in which both may be compared.
 
For reactive scheduling, a modified version of Particon's ISAAC was used. ISAAC is a software, used for energy unit aggregation and planning, based on the principles of controlled self-organization and regulated autonomy ~\parencite{niesse2016isaac}. Here, ISAAC is used to coordinate the negotiations of the agents. 

Within ISAAC, \gls{COHDA} is used to optimize DER scheduling to a given target. Agents exchange information regarding their independently working algorithms to determine the optimal.

The \gls{LPEP}++ was additionally integrated into ISAAC, which allows to start negotiations with both systems. \\

In the scenario, six agents are considered. For each MAS, 100 negotiations were computed. 
The comparison includes the number of messages exchanged, the size of the messages and the duration of the negotiation until the convergence. Results are collected in \Cref{results}. 

\begin{table}
\caption{Simulation Results (Average/Standard Deviation}
\resizebox{\linewidth}{!}{%
\label{results}
\begin{tabular}{lrr}
\toprule
 & \multicolumn{2} {c} {MAS} \\ \cmidrule{2-3}
 & COHDA & LPEP++ \\
\midrule 
Number of messages & 69.228 / 12.87 & 53.5 / 15.746 \\
Message sizes in Byte & 2964.299 / 363.235 &  1650.69 / 102.57  \\
Negotiation period in seconds & 0.0185 / 0.0048 & 0.0180 / 0.023 \\
\bottomrule
\end{tabular}}
\end{table}
The standard deviation of the message size in cohda is due to the fact that the agents send along the system state. This contains the current solution candidate. As the negotiation progresses, the number of agents included in the solution candidate increases, since the agent only knows part of the network at the beginning of the negotiation.
Furthermore, the standard deviation of the negotiation period of the \gls{LPEP} is due to the implementation of the timer.
The \gls{LPEP} meets the advantages such as fast convergence provided by \gls{COHDA}, while still guaranteeing the optimal solution to be found.\\
\subsection{LPEP++ with communication delays}
To state the robustness of the \gls{LPEP}, a communication scenario in the 
Python Library NetworkX was implemented. NetworkX provides functionality for analysing networks and graphs.
A wireless Network was created, each of the six agents was placed at a single node. Between the nodes, three access points were placed. The scenario was used to determine communication delays between two agents. The delay between each agent and its closest access point was chosen randomly between 0.01 and 0.1 seconds. To determine the delay between certain agents, the shortest path according to Dijkstra's algorithm was calculated. When exchaning information, the agent's messages are delayed by the computed time according to the
ICT scenario. 
Results show that with different possible topologies of the agent, the \gls{LPEP} always finds the problems solution.   

\noindent 
\begin{table}
\caption{Simulation Results Delayed LPEP++ (Average/Standard Deviation}
\resizebox{\linewidth}{!}{%
\label{lpep}
\begin{tabular}{lrr}
\toprule
 & \multicolumn{2} {c} {MAS} \\ \cmidrule{2-3}
 & LPEP++ & LPEP++ Delayed \\
\midrule 
Number of messages & 53.5 / 15.746 & 41 / 11.99 \\
Message sizes in Byte &  1650.69 / 102.57 & 2004.69 / 202.57 \\
Negotiation period in seconds & 0.0180 / 0.023 & 0.046 / 0.016 \\
\bottomrule
\end{tabular}}
\end{table}

\section{Conclusion and Future Work}
\label{sec:conclusion}
The protocol extension includes the additional calculation of the applicable power as soon as the disequilibrium cannot be completely dissolved. In addition, the possibility of rescheduling longer-term planning intervals such as entire schedule was demonstrated.  
The modifications and extensions of the protocol lead to advantages in certain use cases and also enable its use in other cases such as the replanning of an entire schedule table, e.g. during intraday planning. 

The solver modification includes the division of the atoms by using the interval boundaries instead of the gcd. This results in a lower number of variables which provides more effiency.

The results show that the protocol finds a solution under acceptable time with acceptable message sizes and can keep up with the heuristic \gls{COHDA} in aspects such as fast convergence, but beyond that it even provides the guarantee of solution completeness which \gls{COHDA} does not.

To simulate the communication infrastructure, delays were determined using a wireless network scenario. It was shown that the protocol extension performs robustly even under non-optimal ICT simulations. The solution is guaranteed to be found even with these communication delays. 

To underpin the robustness of the protocol under poor communication conditions, a communication scenario is to be implemented in OMNET++, which will be linked to the agent implementation. This allows the protocol to be tested under high traffic and packet losses.

\section{Acknowledgements}

The authors would like to thank Max Kronberg and Hendrik Brockmeyer of be.storaged~GmbH for their
contributions towards the protocol extension, Marvin Nebel-Wenner for his valuable help with integrating the LPEP++ into ISAAC, Sebastian Lehnhoff for his counsel and support. This research has been funded by the Federal Ministry for
Economic Affairs and Energy of Germany in the project LarGo! (0350012A).
\balance
\printbibliography

\end{document}

%% file: glossary.inc.tex
\newacronym[shortplural={CPS}]{CPS}{CPS}{cyber-physical system}
\newacronym{AI}{AI}{Artificial Intelligence}
\newacronym{AL}{AL}{Adversarial Learning}
\newacronym{ANN}{ANN}{Artificial Neural Network}
\newacronym{ARL}{ARL}{Adversarial Resilience Learning}
\newacronym{GAN}{GAN}{Generative Adversarial Network}
\newacronym{GPGPU}{GPGPU}{General-Purpose Graphics Processing Unit}
\newacronym{DL}{DL}{Deep Learning}
\newacronym{GRU}{GRU}{Gated Recurrent Unit}
\newacronym{ICT}{ICT}{Information and Communication Technology}
\newacronym{LSTM}{LSTM}{Long-Short Term Memory}
\newacronym{AEG}{AEG}{Attack Execution Graph}
\newacronym{ML}{ML}{Machine Learning}
\newacronym{PoC}{PoC}{Proof-of-Concept}
\newacronym{RL}{RL}{Reinforcement Learning}
\newacronym{DNN}{DNN}{Deep Neural Network}
\newacronym{RNN}{RNN}{Recurrent Neural Network}
\newacronym{CNN}{CNN}{Convolutional Neural Network}
\newacronym{SIMD}{SIMD}{Single Instruction, Multiple Data}
\newacronym{DoE}{DoE}{design of experiments}
\newacronym{DNC}{DNC}{Differentiable Neural Computer}
\newacronym{PV}{PV}{Photovoltaic}
\newacronym{STL}{STL}{Signal Temporal Logic}
\newacronym{MTL}{MTL}{Metric Temporal Logic}
\newacronym{MITL}{MITL}{Metric Interval Temporal Logic}
\newacronym{AV}{AV}{Autonomous Vehicle}
\newacronym[shortplural={MAS}]{MAS}{MAS}{Multi Agent System}
\newacronym{RTU}{RTU}{Remote Terminal Unit}
\newacronym{TCP}{TCP}{Transmission Control Protocol}
\newacronym{LPEP}{LPEP}{Lightweight Power Exchange Protocol}
\newacronym{FMI}{FMI}{Functional Mock-Up Interface}
\newacronym{XML}{XML}{eXtensible Markup Language}
\newacronym{COHDA}{COHDA}{Combinatorial Optimization Heuristic for Distributed Agents}
\newacronym{VPP}{VPP}{Virtual Power Plant}
\newacronym{TVL}{TVL}{Ternary Vector List}

\newglossaryentry{resilience}{
  name=resilience,
  description={The ability of a system to withstand unforseen, rare and
  potentially catastrophic events and to self-heal or improve in
  reaction to these events. Ideally resilience is increasing monotonously.}
}